# Parallel Statistical Model Checking for Safety Verification in Smart Grids

Toni Mancini, Federico Mari, Igor Melatti,
Ivano Salvo, and Enrico Tronci
Computer Science Department,
Sapienza University of Rome, Italy

Jorn Klaas Gruber, Barry Hayes,
and Milan Prodanovic
Instituto IMDEA Energía, Spain

Lars Elmegaard
SEAS-NVE, Denmark

*Abstract*—By using small computing devices deployed at user premises, Autonomous Demand Response (ADR) adapts users electricity consumption to given time-dependent electricity tariffs. This allows end-users to save on their electricity bill and Distribution System Operators to optimise (through suitable time-dependent tariffs) management of the electric grid by avoiding *demand peaks*.

Unfortunately, even with ADR, users power consumption may deviate from the expected (minimum cost) one, *e.g.*, because ADR devices fail to correctly forecast energy needs at user premises. As a result, the *aggregated power demand* may present undesirable peaks.

In this paper we address such a problem by presenting methods and a software tool (APD-Analyser) implementing them, enabling Distribution System Operators to effectively verify that a given time-dependent electricity tariff achieves the desired goals even when end-users deviate from their expected behaviour.

We show feasibility of the proposed approach through a realistic scenario from a medium voltage Danish distribution network.

## I. INTRODUCTION

One of the most challenging problems in the management of modern electric smart grids is to find a trade-off between two conflicting goals. On one hand both the energy retailer, who sells energy, and the Distribution System Operator (DSO), who manages the Electric Distribution Network (EDN), want to sell as much energy as possible, without forcing end-users to cut their power demands. On the other hand, if all users require power at the same time (*peak time*), this may result in higher costs for the DSO (due to, *e.g.*, the need of suddenly activating additional power plants to face such peaks), and possible service and/or infrastructure disruption (due to, *e.g.*, feeder damage or reduced life-time [1]). In fact, the maximum power that can be safely provided by an EDN (*i.e.*, without stressing the infrastructure devices) is *significantly less* than the power that would be needed in the most demanding scenario, when all users request their maximum allowed power simultaneously [2].

Demand Side Management (DSM) [3] approaches are used to achieve the objective of reducing/avoiding the presence of dangerous peaks in the Aggregated Power Demand (APD) (thus performing what is called *peak shaving*). An increasingly popular approach to DSM is Demand Response (DR) (see, *e.g.*, [4] and references therein). In a DR setting, *Inclining Block Rate* or *time-dependent tariffs* are computed by the DSO and proposed to the users, with the goal of *indirectly steering* their individual power demands in such a way that the APD satisfies EDN constraints.

Classically, DR has been implemented by directly including such tariffs in the user contracts. However, this approach poses two main problems: 1) tariffs explicitly included in the user contracts are long-term and *static*, as users must be able to get used to and comfortably follow them; 2) such tariffs might not be enough to sufficiently diversify the individual user power demands, thus simply resulting in peak rebounds (see, *e.g.*, [5]), where the critical peaks of the APD at the substation level are not removed, but just shifted in time.

### A. Motivations

Recent advancements in energy storage technology and electricity usage control algorithms enabled innovative approaches to DR (see, *e.g.*, [6], [7], [8], [9]) where small computing devices (in the following: Home Energy Controllers, HECs) are deployed at user premises to *autonomously* shift high loads (*e.g.*, charging of Plug-in Electric Vehicles, or heating), control local electricity generators (*e.g.*, Photovoltaic panels) and storage (*e.g.*, batteries), compatibly with user actual behaviour and device capabilities (*e.g.*, capacities and power rates of installed batteries), while reacting to electricity price changes. Their ultimate goal is to minimise the user electricity bills.

In such *Autonomous Demand Response (ADR)* settings, the users are freed from the burden to manually adapting their electricity consumption behaviours, as reactions to price changes are handled automatically. All this allows DSOs to propose time-dependent tariffs envisioning *high-frequency* fluctuations (*e.g.*, a new price every 15 minutes) announced with short notice (*e.g.*, 24 hours in advance), exploiting, *e.g.*, quotes of the day-ahead market (see, *e.g.*, [10]). Also, such *price policies* can be diversified among users connected to the same EDN substation (under non-discrimination requirements, see *e.g.*, [4]) in order to induce their HECs to react in complementary ways, ultimately producing an APD compatible with the grid safety constraints. To this end, designed price policies must take into consideration suitable short-term forecasts of each user power demand (which are typically based on historical demand plus, possibly, other data, see, *e.g.*, [11]).

In this paper we focus on such ADR scenarios. The design of *short-term*, *individualised* price policies based on single user power demand forecasts and envisioning *highly fluctuating* prices poses several computational challenges for the DSO. In previous work [4] we proposed global optimisation algorithmic methods to support the price policy design phase, building on mature technologies widely used in many diverse application domains (see, *e.g.*, [12], [13], [14], [15]). Anyway, although global optimisation guarantees to return a set of policies (if one such set exists) that would satisfy all the requirements, in practice the grid safety is still not guaranteed. This is because, due to the limits of HEC capabilities, the actual user power demands may *not* follow the power profiles expected by the DSO as a result of the proposed price policies.

Although the DSO can perform a *worst case analysis* (as in, *e.g.*, [16]) of the impact of the envisioned set of price policies on the EDN safety, such an analysis could be too pessimistic and prevent the DSO from fully exploiting the opportunities stemming from highly fluctuating tariffs and its own investments on the EDN.

*B. Contributions*

In this paper we propose *Aggregated Power Demand Analyser (APD-Analyser)*, a High Performance Computing (HPC)-based software system aimed at supporting DSOs in analysing the possible aggregated effects, on the EDN substation, of individualised price policies for electricity end-users, designed for highly dynamic ADR schemes.

In particular, APD-Analyser allows DSOs to compute a portfolio of Key Performance Indicators (KPIs) over the *probability distribution of the APD*, subject to probabilistic deviations of end-users from their predicted behaviours under the given price policies.

APD-Analyser can then act both as a *Decision Support System* allowing *what-if analyses*, and as an *automated verifier* for the safety of an envisioned set of price policies. In particular, APD-Analyser enables safety assessment of the price policies computed through automated methods (*e.g.*, [4] and references therein).

## II. Aggregated Power Demand Analyser

In this section we describe the input required and the output provided by APD-Analyser. In the following, we denote with $\mathbb{R}$, $\mathbb{R}^+$, $\mathbb{N}$ the sets of, resp., all real, strictly positive real, and non-negative integer numbers.

We assume that the period for which we want to compute the APD probability distribution (*e.g.*, the next day) is split into a set $\mathcal{T}$ of time slots of equal duration (*e.g.*, 15 minutes).

Given $\mathcal{T}$, a *power profile* is a function $p : \mathcal{T} \to \mathbb{R}$. For each $t \in \mathcal{T}$, $p(t)$ denotes a power value (in kW). We use such functions to model both the users' Expected Power Profiles (EPPs) (in this case $p(t)$ gives the average power demanded by a given user during time slot $t$) and the substation safety threshold (in this case $p(t)$ gives the maximum average power that the substation can provide during time slot $t$ in safe conditions).

*A. Inputs*

The system requires the DSO to provide various inputs and configuration options, the most important of which are briefly outlined in the following.

*1) EDN substation end-users:* The finite set $\mathcal{U}$ of end-users (households, commercial buildings, etc.) connected to the EDN substation of interest.

*2) Analysis time period:* The future period of interest, as a set of contiguous time slots $\mathcal{T}$ of the same duration (*e.g.*, 15-minute time slots spanning the next day).

*3) Expected Power Profiles (EPPs):* The EPPs of all end-users in $\mathcal{U}$, defined under the (optimistic) assumption that end-users will adhere to the price policies under analysis, as predicted by the DSO. Such EPPs may be either computed by some grid-level service [4] or forecasted using, *e.g.*, users historical data or other information (see, *e.g.*, [11]). EPPs are defined in terms of a power profile $p_u$ over $\mathcal{T}$ for each $u \in \mathcal{U}$. Note that EPPs functions can take *negative* power values, to account for, *e.g.*, power generation capabilities at end-user premises, for example Photovoltaic installations.

*4) Users deviation model:* A *probabilistic model* of end-user deviations wrt. their expected behaviours, given in terms of a function $\text{dev}_u : D_u \to [0, 1]$ for each $u \in \mathcal{U}$, where $D_u$ is a bounded interval of real values defining possible deviations of $u$ wrt. EPP $p_u$. For each user $u \in \mathcal{U}$, $\text{dev}_u$ defines a probability density function (over bounded support $D_u$), hence $\int_{D_u} \text{dev}_u(x) \mathrm{d}x = 1$. For each $u \in \mathcal{U}$ and $a, b \in D_u$ ($a < b$), $\int_a^b \text{dev}_u(x) \mathrm{d}x$ defines the probability that $u$ will deviate from their EPP by a factor between $a$ and $b$, *i.e.*, that in any time slot $t \in \mathcal{T}$, the actual power demand of $u$ is within $[(1+a)p_u(t), (1+b)p_u(t)]$. User deviation probabilities can be retrieved from historical data, *e.g.*, by comparing past users' EPPs to their actual behaviours. Also, to overcome the intrinsic uncertainty of users behaviour predictability, sets $D_u$ of users' possible deviations can be rather generous (*e.g.*, in our case study of Section VI we allowed, for *all* users, deviations as high as $\pm 40\%$ wrt. their EPPs).

*5) Substation safety requirements:* The substation safety thresholds during $\mathcal{T}$, in terms of a power profile $p_s : \mathcal{T} \to \mathbb{R}$. To keep the substation line (and hence the EDN) safe, the APD must be less than or equal to $p_s(t)$ in any time slot $t \in \mathcal{T}$.

*6) KPIs:* A finite set $\mathcal{K}$ of KPIs defined over the probability distribution of the APD of the EDN substation, for different values of the substation safety requirements $p_s$.

Defining KPIs as general functions of the APD probability distributions is a flexible way to take into account DSO requirements. As an example, a KPI might define the probability that the APD exceeds the substation safety power threshold when such a threshold is at most $v$. Other typical KPIs would rank APD probability distributions according to their similarity to desired shapes (defined, *e.g.*, starting from the electricity quotations in the day-ahead market).

*7) Parameters:* Real-valued parameters $\varepsilon, \delta \in (0, 1)$ and $\gamma \in \mathbb{R}^+$ defining, resp., *tolerance* and *statistical confidence* of the computed estimated values of the APD probability distribution (see Definition 1) and its *discretisation step* (see Section II-B).

*B. Outputs*

The goal of APD-Analyser is to compute values for all the given KPIs under the given inputs. As KPIs are functions of the APD probability distributions for different values of the substation safety requirements $p_s$, the *core* computational task of APD-Analyser is to compute, for each value $v$ of the substation safety requirements, the probability distribution of the APD when the substation can safely provide power $v$. Such a distribution in turn depends on the EPPs of all users and their probabilistic deviation models.

Unfortunately, an exact computation of the required APD probability distributions is prohibitive, because of its exponential dependence in the number of users and time-slots (see Section VI-C for an example). Hence, APD-Analyser computes so called $(\varepsilon, \delta)$-approximations (Definition 1) of the probabilities of interest.

Formally, let $[w_{\min}, w_{\max}]$ be the interval enclosing all the possible values that the APD can take, given the users EPPs and their deviation models (defining a bounded range of possible deviations $D_u$ for each $u \in \mathcal{U}$): $w_{\min} = \min_{t \in \mathcal{T}} \left[ \sum_{u \in \mathcal{U}} (1 + \min(D_u)) p_u(t) \right]$ and $w_{\max} = \max_{t \in \mathcal{T}} \left[ \sum_{u \in \mathcal{U}} (1 + \max(D_u)) p_u(t) \right]$. Let $W$ be a covering of $[w_{\min}, w_{\max}]$ into a finite set of disjoint and contiguous intervals of width $\gamma$ (which we call *power slots*), i.e., $W = \{[x, x + \gamma) \mid x < w_{\max}, x = w_{\min} + k\gamma, k \in \mathbb{N}\}$.

Let $V_s = \{v \mid p_s(t) = v, t \in \mathcal{T}\}$ be the set of all the (distinct) values for the power safely provided by the substation in some time slot. Also, for each $v \in V_s$ and $w \in W$, let $\Psi_v(w)$ be the (exact but unknown) probability that the APD takes a value within power slot $w$ when the substation safety threshold is $v$.

Then, APD-Analyser returns a set of functions $\{\tilde{\Psi}_v : W \to \mathbb{R} \cup \{\bot\} \mid v \in V_s\}$ (with $\bot$ being a distinguished value) such that, for each $v \in V_s$, $\tilde{\Psi}_v$ has the following properties (see Theorem 1):

- if $\tilde{\Psi}_v(w) = \bot$ for a power slot $w \in W$, then, with statistical confidence at least $1 - \delta$, the probability $\Psi_v(w)$ is less than $\varepsilon$;
- for all the other power slots $w$, with statistical confidence at least $1 - \delta$, value $\tilde{\Psi}_v(w)$ is within $(1 \pm \varepsilon)\Psi_v(w)$.

## III. RANDOM VARIABLE MEAN $(\varepsilon, \delta)$-APPROXIMATION

Many applications, as our APD-Analyser, require the computation of the mean value $\mu_Z$ for a 0/1 random variable $Z$. When, as in our case, an exact computation is infeasible, Monte Carlo methods are often used to compute an approximation of this quantity [17], [18].

The main idea is to consider $N$ independent and identically distributed (iid) random variables $Z_1, \ldots, Z_N$ identical to $Z$ (having mean $\mu_Z$), or, equivalently, $N$ independent realisations (samples) of the random variable $Z$, and to take $\tilde{\mu}_Z = \frac{(Z_1 + \cdots + Z_N)}{N}$ as an approximation of $\mu_Z$.

In order to have a formally guaranteed upper bound for the error of this approximated computation, we exploit the notion of $(\varepsilon, \delta)$-*approximation* of the mean value of a random variable (Definition 1).

**Definition 1** (($\varepsilon, \delta$)-approximation, [17]). *Let $\mu_Z$ be the mean value for a 0/1 random variable $Z$ and $Z_1, \ldots, Z_N$ be $N$ independent realisations (samples) of $Z$. Given $\varepsilon$ (tolerance) and $\delta$ (statistical confidence) in $(0, 1)$, we say that $\tilde{\mu}_Z = \frac{(Z_1 + \cdots + Z_N)}{N}$ is an $(\varepsilon, \delta)$-approximation of $\mu_Z$ if:* $\Pr\left[(1 - \varepsilon)\mu_Z \leq \tilde{\mu}_Z \leq (1 + \varepsilon)\mu_Z\right] \geq 1 - \delta$.

## IV. MONTE CARLO–BASED COMPUTATION OF APD PROBABILITY $(\varepsilon, \delta)$-APPROXIMATIONS

The APD-Analyser Monte Carlo–based algorithm computes, for each distinct value $v \in V_s$ of the substation safety threshold in $\mathcal{T}$ and for each power slot $w \in W$, an $(\varepsilon, \delta)$-approximation $\tilde{\Psi}_v(w)$ of the probability that the APD takes a value within $w$ when the substation safety threshold is $v$.

Given the huge number of samples to be generated and evaluated, a naive implementation of a Monte Carlo algorithm to estimate an $(\varepsilon, \delta)$-approximation $\tilde{\Psi}_v(w)$ of $\Psi_v(w)$ for each power slot $w \in W$ would be infeasible.

To this end, for each value $v \in V_s$, our algorithm generates iid random samples of the APD, starting from the EPP and the deviation model of each $u \in \mathcal{U}$. Each generated APD sample $w \in W$ is used to define the realisation of a set of $|W|$ 0/1 random variables $\{Z^i \mid i \in W\}$, one per power slot, where $Z^w$ is 1 and all the others are 0. Given a set of $N$ iid samples $w_1, \ldots w_N$ (in $W$) generated as above, we have that, for each power slot $i \in W$, the 0/1 variables $\{Z_1^i, \ldots Z_N^i\}$ are iid. This results in our algorithm effectively computing $(\varepsilon, \delta)$-approximations $\tilde{\Psi}_v(w)$ of $\Psi_v(w)$ for each power slot $w \in W$ using a *single* set of samples in $W$.

APD-Analyser combines two Statistical Model Checking approaches. The first approach, based on the algorithm in [17], restricted to 0/1 random variables, decides when to stop sampling. The second approach, important to guarantee termination on zero/low-probability areas of the computed distribution, relies on Statistical Hypothesis Testing (along the lines of [19], [20], [21]).

Our method uses the condition in [17] to decide whether $N_i$ is sufficiently large to guarantee that $\frac{1}{N_i} \sum_{j=1}^{N_i} Z_j^i$ is a $(\varepsilon, \delta)$-approximation $\tilde{\Psi}_v(i)$ of the probability $\Psi_v(i)$ that the APD takes a value within power slot $i$ when the substation safety threshold is $v$. At the same time, if after a sufficiently large number of samples $M_i$, $\sum_{j=1}^{M_i} Z_j^i$ is still 0 for some power slot $i$, the algorithm concludes that, with confidence at least $1 - \delta$, $\tilde{\Psi}_v(i) < \varepsilon$.

## V. A PARALLEL ALGORITHM FOR APD-ANALYSER

Our APD-Analyser algorithm has been designed as to be massively parallelised on a HPC infrastructure.

To this end, the computation is split in many Workers (Section V-A) and one Orchestrator (Section V-B).

*A. Workers*

Each worker runs in parallel upon requests from the Orchestrator. To reduce the overhead due to network communication,

each invocation to a worker triggers the generation of a high number $c$ of samples. Each sample of the APD (when the substation safety threshold is $v$) is generated as follows:

1) A time slot $t \in \mathcal{T}_v = \{t | t \in \mathcal{T}, p_s(t) = v\}$ is extracted uniformly at random (as time slots have equal duration).
2) For each user $u \in \mathcal{U}$, a deviation $d_u \in D_u$ is extracted according to the deviation model of $u$ (*i.e.*, using probability density function $\text{dev}_u$).
3) The APD value in time-slot $t$ resulting from the extracted user deviations is computed as $A = \sum_{u \in \mathcal{U}} (1 + d_u) p_u(t)$ and associated to its power slot $w = [a, a + \gamma) \in W$ (*i.e.*, $w$ is such that $a \leq A < a + \gamma$).

At the end of each sampling task, the $c$ generated samples (power slots) are sent back to the Orchestrator.

### B. Orchestrator

The Orchestrator runs the two Statistical Model Checking algorithms by asking workers to generate new samples and by properly aggregating the results. The Orchestrator runs a *round* of the above for each substation power value $v \in V_s$. For each round (*i.e.*, for each $v \in V_s$) and for each power slot $w \in W$, the computation state of $\tilde{\Psi}_v(w)$ may be in one of two *phases*, which evolve *independently*. Also, for each power slot $w \in W$, the following *null hypothesis* $H_0^w$ is formulated: "$\Psi_v(w) \geq \varepsilon$".

*Phase 1:* 0/1 samples are generated for $w$ (starting from the samples over $W$ as described earlier) until one of two conditions arises:

(a) $\Upsilon_1$ samples are 1 (where $\Upsilon_1$ is a function of $\varepsilon$ and $\delta$ as defined in [17]).
(b) The first $\lceil \ln(\delta) / \ln(1 - \varepsilon) \rceil$ samples are all 0.

Let $N_1^w$ be the number of samples generated for $w$. In case (a), a first estimation of $\Psi_v(w)$ is computed as ratio $\frac{\Upsilon_1}{N_1^w}$, and the computation proceeds with "Phase 2".

In case (b), the null hypothesis $H_0^w$ is *rejected*, thus declaring that $\Psi_v(w) < \varepsilon$. This Statistical Hypothesis Rejection procedure is known to be a one-sided error decision algorithm, as it may introduce a *type-I error* (*i.e.*, the algorithm might reject $H_0^w$ when it actually holds). However, from [19] it follows that the probability of such an error, $\Pr\left[Z_1^w = \cdots Z_{N_1}^w = 0 \mid H_0^w\right] < \delta$. This equivalently tells us that $H_0^w$ is rejected with confidence at least $1 - \delta$. In case (b) value for $\tilde{\Psi}_v(w)$ is set to special value $\bot$ and the computation for power slot $w$ moves to state "Done".

*Phase 2:* For any power slot $w$ for which the computation state of $\tilde{\Psi}_v(w)$ is "Phase 2", the Orchestrator waits until a number of samples greater than $N_2^w$ is collected. Value for $N_2^w$ is computed during "Phase 1", and depends on the first estimate of $\tilde{\Psi}_v(w)$ computed therein. Value for $N_2^w$ is needed in Phase 2 to compute the number of *additional* samples $N_3^w$ that must be generated to compute the final $(\varepsilon, \delta)$-approximation of $\tilde{\Psi}_v(w)$. When such additional $N_3^w$ 0/1 samples $\{Z_1^w, \ldots, Z_{N_3^w}^w\}$ are generated by the workers, the final $\tilde{\Psi}_v(w)$ is set to $\frac{1}{N_3^w} \sum_{i=1}^{N_3^w} Z_i^w$ and the computation for $w$ is set to state "Done".

The Orchestrator terminates the round for $v \in V_s$ when the computation process of $\tilde{\Psi}_v(w)$ is in state "Done" for *all* power slots $w \in W$.

### C. Algorithm correctness

The following result holds (proof omitted for lack of space).

**Theorem 1.** *For each $v \in V_s$ and each $w \in W$, let $\tilde{\Psi}_v(w)$ be the (unknown) probability that the Aggregated Power Demand (APD) is within power slot $w$ in a time slot extracted from $\mathcal{T}_v = \{t | t \in \mathcal{T}, p_s(t) = v\}$ uniformly at random, when each user $u \in \mathcal{U}$ independently deviates from the predicted collaborative profile $p_u$ according to the deviation model $\text{dev}_u$.*

*Value $\tilde{\Psi}_v(w)$ computed by the algorithm above is such that:*
*1) If $\tilde{\Psi}_v(w) = \bot$, then $\Psi_v(w) < \varepsilon$ with confidence $\geq 1 - \delta$.*
*2) Otherwise, $\tilde{\Psi}_v(w)$ is a $(\varepsilon, \delta)$-approximation of $\Psi_v(w)$.*

## VI. EXPERIMENTAL RESULTS

In this section we present experiments aiming at evaluating our C implementation of APD-Analyser in computing an $(\varepsilon, \delta)$-approximation of APD probability values and the computational scalability of our parallel algorithm.

### A. Case study

For our case study, we used real electricity usage data from 130 residential households (set $\mathcal{U}$) connected to a substation of a medium voltage distribution network in Denmark (data courtesy of SEAS-NVE). EPPs for all users are defined with a granularity of 60 minutes. The (constant) safety threshold for APD is 400 kW, as this is the substation nominal power.

We adopted the same deviation model for all users $u \in \mathcal{U}$, which defines the range of possible deviations as $D_u = [-0.4, 0.4]$ and density function $\text{dev}_u$ defined as: $\text{dev}_u(x) = 0.4\delta(x) + 0.2\delta(x - 0.2) + 0.2\delta(x + 0.2) + 0.1\delta(x - 0.4) + 0.1\delta(x + 0.4)$ (where $\delta(x)$ is the Dirac delta function). That is, users can deviate from their EPPs with non-zero probability in 5 possible ways, although very generously (as the largest possible deviations are as large as $\pm 40\%$ and occur with probability as high as $10\%$). Using a finite number of user deviations was dictated by the structure of our data and specifications from the DSO. However, we stress that such a choice has no impact on the performance of our tool, which uses standard libraries for the efficient sampling from continuous distributions.

### B. Experimental setting

We ran our APD-Analyser on a HPC infrastructure, allocating up to 81 nodes (1 Orchestrator and up to 80 workers). Each process ran on a node consisting of an Intel(R) Xeon(R) CPU with a frequency varying within 2.27–2.83 GHz, and had access to 1 GB of RAM.

### C. Computation of APD probability distributions

In order to assess effectiveness of our parallel algorithm to compute $(\varepsilon, \delta)$-approximations of the APD probability distribution values, we considered a very challenging scenario from our case study.

In particular, using data from all the 130 users in our dataset, we computed $(\varepsilon, \delta)$-approximations of the APD probability distribution values for each month of the year (time slot duration: 1 day). As each of the 130 users can independently deviate from their EPPs by 5 possible factors for each time slot, the overall allowed number of combinations of user deviations is approximately $5^{30 \times 130}$. This makes an exhaustive exploration of the space of user deviations (hence an exact computation of the APD distribution) clearly impossible. We set $\varepsilon$, $\delta$ and $\gamma$ to small values, namely $\varepsilon = \delta = 5\%$ ($5 \times 10^{-2}$), $\gamma = 20$ kW (*i.e.*, 5% of the substation nominal power). Finally, to reduce network overhead, we set each worker as to generate $c = 2 \times 10^4$ samples per single Orchestrator request.

Figure 1 shows the APD probability distribution computed by our algorithm for each month under these settings. Each of the 12 computations took between 4782 seconds and 6448 seconds (average: 5297 seconds, *i.e.*, 1 hour, 28 minutes and 17 seconds) using 80 workers, and involved the generation of a number of samples ranging from $\approx 1.6 \times 10^9$ to $\approx 2.2 \times 10^9$ (average: $\approx 1.8 \times 10^9$ samples).

With current Infrastructure as a Service (IaaS) pricing schemes, the cost for such a computation would be of around Eur 2.80, *i.e.*, as small as Eur 0.02 per user per day.[1]

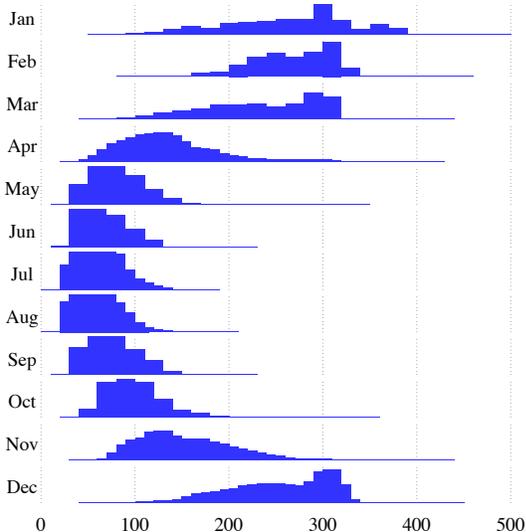

Figure 1: Monthly APD (in kW) probability distributions, as computed by our algorithm ($\varepsilon = \delta = 5\%$, $\gamma = 20$ kW) on our case study scenarios.

### D. Scalability analysis

In order to evaluate the scalability of our parallel algorithm, we ran additional experiments by varying $k$, the number of used parallel workers/computational nodes ($k \in \{1, 20, 40, 60, 80\}$), and measuring the number of samples per second generated (and processed by the Orchestrator) in each configuration. For each value of $k$, we computed the *speedup* and *efficiency* of our parallel algorithm as $\frac{s_k}{s_1}$ and

---

[1] June 2018 prices for 80 Google g1-small nodes for 90 minutes, see https://cloud.google.com/products/calculator

---

$\frac{s_k}{s_1 \times k}$ resp., where $s_k$ and $s_1$ are the number of *samples per second* received and aggregated by the Orchestrator when using, respectively, $k$ and 1 workers (see, *e.g.*, [22]).

Table I shows that our parallel algorithm scales very well with the number of workers (computational nodes), with efficiency values stabilising at around $70\%$, a typical value in cluster environments.

The short computation time needed by APD-Analyser clearly enables its *online* usage, in order to compute the KPI values for *multiple* candidate price policies, hence assisting the DSO in price policy *design*. To this end, it is worth noting that the stable efficiency of our parallel algorithm (as shown in Table I) allows DSOs to compress computation times by increasing the number of computational nodes used (with no substantial additional IaaS cost). For example, using 501 computational cores (*i.e.*, 1 Orchestrator and 500 Workers) and assuming an efficiency of 70%, the expected computation time would have been of just 15 minutes (for an IaaS cost of around Eur 2.90 per day, *i.e.*, Eur 0.02 per user per day, see Footnote 1).

| # workers | samples/sec | speedup | efficiency |
|---|---|---|---|
| 1 | 5924.89 | 1× | 100% |
| 20 | 79275.028 | 13.38× | 66.90% |
| 40 | 162578.98 | 27.44× | 68.60% |
| 60 | 257791.96 | 43.51× | 72.52% |
| 80 | 335823.24 | 56.68× | 70.85% |

Table I: APD-Analyser parallel algorithm scalability.

## VII. RELATED WORK

In statistical model checking approaches (see, *e.g.*, [23], [24], [25], [26], [27], [28], [29], [19], [17]), scenarios are sampled until a certain degree of confidence in the estimated probability is achieved. APD-Analyser is based on such techniques, and employs a parallel version of the algorithms in [19], [17], also adapting them to the smart grid context. There exist other works adapting statistical model checking techniques to smart grids, see, *e.g.*, [30], [31], which however are based on sequential (non-parallel) algorithms. In a smart grid context, usage of *exhaustive* (instead of statistical) methods for safety analysis has been studied, *e.g.*, in [32], [33], [34]. All such works are plagued by the *state explosion* problem. Since APD-Analyser aims at evaluating safety of a large system (a substation connecting hundreds of end-users), exhaustive or non-massively-parallel approaches *cannot* be used.

Monte Carlo–based methods aiming at evaluating the average behaviour of the system under analysis are widely used in *performance evaluation* where, typically, the average case is the most interesting one. However such methods, used in a smart grid context for example in [35], [36], are not very well suited for safety analysis, since they cannot address rare (low probability) events and, most importantly, do not provide any lower bound for the statistical confidence achieved.

Performance evaluation can also be done using suitable benchmarks. This is often quite interesting since it guarantees that the system is evaluated on relevant case studies (see,

*e.g.*, [37], [38]). However, this approach cannot be used for safety verification, since it cannot provide an upper bound to the probability that a potentially unsafe scenario has not been considered in the analysis.

## VIII. Conclusions

In this paper we presented APD-Analyser, a HPC-based software system aimed at supporting DSOs in analysing the possible aggregated effects, on an EDN substation, of new envisioned, possibly highly-fluctuating and individualised price policies for electricity end-users. APD-Analyser uses information about DSO-expected user behaviours under the new policies as well as probabilistic models of user deviations from such expected behaviours, thus acting both as a *Decision Support System* (as it allows *what-if analyses*) and as an *automated verifier* for the *safety* of an envisioned set of price policies. In particular, APD-Analyser enables safety assessment of the price policies computed through automated methods as part of highly dynamic ADR schemas.

Feasibility and scalability of the system have been assessed by using realistic scenarios taken from an existing medium voltage Danish distribution network.

*Acknowledgements:* This work was partially supported by the following research projects/grants: Italian Ministry of University & Research (MIUR) grant "Dipartimenti di Eccellenza 2018–2022" (Dept. Computer Science, Sapienza Univ. of Rome); EC FP7 project SmartHG (Energy Demand Aware Open Services for Smart Grid Intelligent Automation, 317761); INdAM "GNCS Project 2018".